# Fluorescence Imaging In Vivo at Wavelengths beyond 1500 nm

*Shuo Diao,[+,a] Dr. Jeffrey L. Blackburn,[+,b] Dr. Guosong Hong,[+,a] Alexander L. Antaris,[a] Dr. Junlei Chang,[c] Dr. Justin Z. Wu,[a] Bo Zhang,[a] Dr. Kai Cheng,[d] Prof. Calvin J. Kuo,[c] Prof. Hongjie Dai*[a]*

[a] Department of Chemistry, Stanford University
Stanford, California 94305 (USA)
E-mail: hdai@stanford.edu

[b] Chemical and Materials Science Center
National Renewable Energy Laboratory
Golden, Colorado 80401 (USA)

[c] Division of Hematology, School of Medicine, Stanford University
Stanford, California 94305 (USA)

[d] Department of Radiology and Bio-X Program, School of Medicine, Stanford University
Stanford, California 94305 (USA)

[+] These authors contribute to the work equally.

**Abstract: Compared to imaging in the visible and near-infrared regions below 900 nm, imaging in the second near-infrared window (NIR-II, 1000—1700 nm) is a promising method for deep-tissue high-resolution optical imaging in vivo mainly due to the reduced scattering of photons traversing through biological tissues. Herein, semiconducting single-walled carbon nanotubes with large diameters were used for in vivo fluorescence imaging in the long-wavelength NIR region (1500—1700 nm, NIR-IIb). With this imaging agent, 3-4 µm wide capillary blood vessels at a depth of about 3 mm could be resolved. Meanwhile, the blood-flow speeds in multiple individual vessels could be**

**mapped simultaneously. Furthermore, NIR-IIb tumor imaging of a live mouse was explored. NIR-IIb imaging can be generalized to a wide range of fluorophores emitting at up to 1700 nm for high-performance in vivo optical imaging.**

Fluorescence-based optical imaging is indispensable to investigating biological systems with high spatial and temporal resolution.[1] However, a formidable challenge to in vivo fluorescence imaging of live animals has been the limited depth of penetration and inability of high-resolution imaging through live tissues owing to both the absorption and scattering of photons. To circumvent this problem, we and others have recently explored in vivo fluorescence imaging in the second near-infrared window (NIR-II, 1000—1700 nm) to benefit from reduced photon scattering and achieve higher imaging resolution deeper in the body than with traditional NIR imaging (NIR-I, 750—900 nm).[2] Nevertheless, the NIR-II fluorophores used thus far mainly emit below about 1400 nm, which is still not optimal from the tissue scattering point of view. Although in vivo fluorescence imaging at longer wavelengths would further reduce scattering, increased water absorption upon approaching the infrared region could diminish the intensity of light passing through biological tissues, which is a valid concern that has deterred in vivo fluorescence imaging in the long wavelength NIR region. Another consideration for the development of in vivo fluorescence imaging in the long NIR window is the lack of biocompatible emitters with sufficient brightness. To date, most NIR-II in vivo fluorescence imaging has been based on high-pressure carbon monoxide conversion (HiPCO) single-walled nanotubes (SWNTs) with a small diameter distribution of 0.7--1.1 nm, emitting band-gap fluorescence in the range of 1000—1400 nm.[2c,g,h,3] We expect that SWNTs with larger diameters, such as SWNTs grown by the laser vaporization method originally developed by Smalley,[4] could enable high-resolution fluorescence imaging in the longer-wavelength region.

The long wavelength NIR-II region near 1600 nm offers a balance of photon-scattering and water-absorption effects, promising to significantly enhance the performance of fluorescence imaging in vivo, by achieving both improved penetration depth and imaging resolution. As shown in the absorption spectrum of water and the extinction spectra of biological tissues (Figure 1a), the near 1600 nm region resides in a local valley between a water vibrational OH stretching overtone absorption peak at approximately 1450 nm and the edge of an increasing water absorption combination band beyond 1700 nm. Thus, the fluorescence imaging window near 1600 nm offers a local minimum in the water absorption spectrum to minimize attenuation of the fluorescence signal caused by water dominant in biological tissues. On the other hand, as photon scattering scales as $\lambda^{-\alpha}$ ($\alpha$=0.2--4 for different tissues),[5] the near 1600 nm window provides the lowest photon scattering in the entire NIR-II window (without water absorption dominance), useful for high-resolution, deep-tissue biological imaging.

We herein report the successful in vivo fluorescence imaging (under 808 nm excitation) in the 1500—1700 nm NIR-II window (referred to as the "NIR-IIb" window). Semiconducting SWNTs with improved fluorescence brightness are chemically enriched from unsorted SWNTs synthesized by laser vaporization. The NIR-IIb imaging window affords an in vivo vascular imaging spatial resolution down to approximately 4 μm at a depth of up to about 3 mm in the mouse hindlimb and brain with intact skull and scalp. Simultaneous single-vessel-resolved blood-flow speed mapping for multiple hindlimb arterial vessels is achieved by video-rate fluorescence imaging in the NIR-IIb region. Furthermore, high-performance NIR-IIb tumor imaging with clear and sharp resolution is explored.

The SWNT NIR-IIb emitters with a diameter range of approximately 0.96--1.24 nm were synthesized by laser vaporization (LV).[6] Changing the furnace temperature during LV synthesis from 950 ℃ to 1125 ℃ changed the SWNT average diameter from approximately 0.9 to 1.4 nm, allowing us to identify the SWNTs grown at 950 ℃ to be the brightest for imaging in the NIR-IIb window (Supporting Information, Figure S1). Compared to the previously widely used HiPCO SWNTs, the LV nanotubes exhibited higher fluorescence in the 1500—1700 nm region owing to the smaller band gaps and the larger average diameter (Figure S2; quantum yield measurements are shown in Figure S3).[4b] The LV SWNTs grown at 950 ℃ were then suspended by sonication and sorted through a chromatography column to enrich the semiconducting SWNTs (see the Method Section in the Supporting Information).[2a,7] Compared to the raw material, the resulting semiconducting LV SWNTs exhibited enhanced $S_{11}/S_{22}$ (ca. 700—1800 nm) peaks and reduced $M_{11}$ peaks (ca. 500—700 nm) in the UV/Vis/NIR absorption spectra (Figure 1c), indicating successful enrichment of the semiconducting LV SWNTs needed for biological imaging with sufficiently bright fluorescence in the NIR-IIb window (Figure 1b,d).

The semiconducting enriched LV SWNTs were stably suspended in aqueous biological solutions with highly biocompatible surface coatings by employing an exchange functionalization method (see the Supporting Information).[2g] We evaluated the wavelength-dependent non-invasive fluorescence imaging of mouse brain vessels at depths of up to 3 mm through the intact scalp and skull of C57Bl/6 mice (after shaving to remove hair) by intravenously injecting indocyanine green (ICG, emitting in the NIR-I region) or a mixture of biocompatible HiPCO and semiconducting LV SWNTs (Figure 2a,c). The images were recorded using a liquid-nitrogen-cooled 2D InGaAs camera (Princeton Instruments, detection range below ca. 1700 nm), which showed quantum efficiencies below 30% at wavelengths above 1600 nm;[8] therefore, most of the NIR-IIb signals recorded in this work were within the 1500—1600 nm region. We measured the signal-to-background ratios (SBRs) by plotting the cross-sectional intensity profiles of the same vessel imaged in the NIR-I (850-900 nm), NIR-II (1000—1700 nm), and NIR-IIb (1500—1700 nm) windows (Figure 2d, f). The SBR obtained by imaging in the NIR-IIb window was found to be higher than for the NIR-I and NIR-II windows (4.50 in NIR-IIb vs. 2.01 in NIR-II and 1.19 in NIR-I), suggesting that in vivo imaging beyond 1500 nm benefits from increased signal-to-background ratios. We also compared the photon scattering effects in the NIR-II and NIR-IIb windows and a previously defined NIR-IIa region (1300—1400 nm; Figure 1a)[2b] by performing phantom studies using both chicken breast (Figure S4) and intralipid (a scattering medium; Figure S5), showing the benefit of minimized photon scattering in the long-wavelength NIR-IIb region for fluorescence imaging with deep penetration.

We also performed high-magnification microscopic vessel imaging of both mouse hindlimb and brain with a pixel size of 2.5 μm in the NIR-IIa and NIR-IIb regions. By measuring the Gaussian-fitted full width at half maximum (FWHM) of the cross-sectional intensity profiles of the features, we found that the apparent widths of the same hindlimb blood vessel imaged at a depth of approximately 3 mm in the NIR-IIa (marked by a green line, Figure 3a) and NIR-IIb (marked by a green line, Figure 3b) were 9.7 μm and 6.3 μm, respectively, suggesting an improved spatial resolution in the longer-wavelength NIR-IIb window (Figure 3c). Also, we evaluated the SBRs of a blood vessel (marked by white lines, Figure 3a,b)imaged in both windows. A higher SBR of 3.85 was obtained in the NIR-IIb than in the NIR-IIa window (1.56) for the same blood vessel (Figure 3d). The smallest vessel resolved by NIR-IIb fluorescence imaging in the mouse hindlimb showed an apparent width (i.e., FWHM) of approximately 3.7 μm at a depth of about 2.6 mm (Figure 3e,f).

We then compared the performance of high-magnification cerebrovascular imaging in the NIR-IIa[2b] and NIR-IIb regions. Many vessels imaged in the NIR-IIb window showed generally improved feature sharpness (Figure 3i; FWHM ca. 5.4 μm in NIR-IIb vs. ca. 9.4 μm in NIR-IIa for the same blood vessel; green lines in Figure 3g,h) and increased SBRs (Figure 3j; 7.44 in NIR-IIb vs. 4.91 in NIR-IIa for the same vessel; white lines in Figure 3g,h). A small vessel with a FWHM width down to about 4 μm was measured in the NIR-IIb region at a depth of approximately 2.8 mm through the intact scalp and skull, achieving micrometer resolution for in vivo brain fluorescence imaging at >2 mm depth (Figure 3k,l). The imaging depth was determined throughout this work by recording the vertical axis travel distance of the microscopic objective with the surface of the skin set to be a depth of zero. Ex vivo mouse tissue imaging was also performed to study the scattering effects in vessel size measurements in the NIR-IIb region (Figure S6).

To perform video-rate fluorescence imaging of blood vessels in the mouse hindlimb in the NIR-IIb region, we injected a 200 μL solution of exchanged semiconducting LV SWNTs into the tail vein of an athymic nude mouse and performed NIR-IIb fluorescence imaging at a speed of 4.6 frames per second (Movie S1). Upon injection, blood flow into the femoral artery was clearly observed by video-rate imaging (Figure 4a). Aside from the main femoral artery, blood flow into numerous smaller, high-order arterial vessel branches was also observed (Figure 4b,c and Movie S1). These smaller vessels were deeper in the body than the femoral artery and were difficult to resolve by imaging in the 1000—1400 nm[2c] or 1500-1700 nm (Figure S7) range using HiPCO SWNTs. The blood-flow speeds in various arterial vessels were quantified from video-rate imaging (Movie S1) by plotting the distance travelled by the signal front versus time (Figure S8),[2c] affording a spatially resolved blood flow map showing blood flow speeds of approximately 22.6 mm $s^{-1}$ in the femoral artery and 1.1 to 4.8 mm $s^{-1}$ in the smaller arterial branches (Figure 4f). To the best of our knowledge, this is the first time that blood flow velocities over a broad range of 1—20 mm $s^{-1}$ in multiple vessels of mice were mapped simultaneously. Blood flow velocimetry over such a broad dynamic range from about 1 mm $s^{-1}$ to >20 mm $s^{-1}$ with spatially resolved individual vessels is hard to achieve with other standard techniques, such as laser Doppler and ultrasonography, which suffer from low spatial resolution, speckle artifacts, and a small dynamic range.[2c,3c,9]

The femoral veins were observed at about 26 s post injection (p.i.; Figure 4e), as these vessel structures became fluorescent in the NIR-IIb region owing to the blood flow completing a systemic circulation cycle in the hindlimb. Principal component analysis (PCA) of the video-rate image frames was performed to group the image pixels with similar time variance into distinct components, clearly identifying the various arterial branches and differentiating them from the venous vessels (Figure 4f).[2c,3a,10]

Lastly, we explored the NIR-IIb emitting SWNTs for in vivo tumor imaging. After intravenously injecting exchanged semiconducting LV SWNTs into a balb/c mouse bearing two subcutaneous 4T1 murine breast tumors, we performed video-rate (4.6 frames per second) NIR-IIb fluorescence imaging of the mouse under 808 nm excitation (Movie S2). Because of the pulmonary circulation, an intensity spike in the lungs (a deep organ in mice) was clearly resolved after injection into the tail vein (Figure 5b,c). Afterwards, NIR-IIb signals in the major organs (such as the kidneys) in the systemic circulation were clearly observed (Figure 5d,e). The vascular structures surrounding the tumors started to appear in a prominent manner at about 3.9 s p.i. (Figure 5e,f). These deep vessels were difficult to clearly resolve by imaging in shorter-wavelength regions owing to scattering by muscles and other tissues atop (Figure S10).[3b] Registration of the inner organs and vessels was evident by applying PCA to the frames of the video-rate movie (Figure 5g). Over the SWNT blood circulation (half-life of about 5.6 h; Figure S11),[11] the nanotubes gradually accumulated within the tumors owing to the enhanced permeability and retention (EPR) effect (Figure 5h).[3b,12] Aside from tumor uptake, the SWNTs were mainly accumulated in the liver and spleen of the reticuloendothelial system (RES) at seven days post injection (Figure S11).

In conclusion, we have shown that the long-wavelength NIR imaging window of 1500—1700 nm is advantageous for in vivo fluorescence imaging as photon scattering is minimized. Water absorption is slightly higher in the NIR-IIb than in the traditional NIR window, but the absorption effects can be overcome with sufficiently bright fluorophores. This is an approach worth taking as scattering problems cannot be easily solved or avoided unless longer-wavelength photons are detected. NIR-IIb imaging offers the lowest light scattering among all the NIR sub-regions examined, affording high spatial resolution, deep tissue penetration, and high signal/background ratios for in vivo biological imaging. The results here are not limited to nanotube fluorophores, and should be generalized to other agents emitting in the region up to 1700 nm, especially for fluorophores with much higher quantum yields, for further enhancing NIR-IIb imaging capabilities.

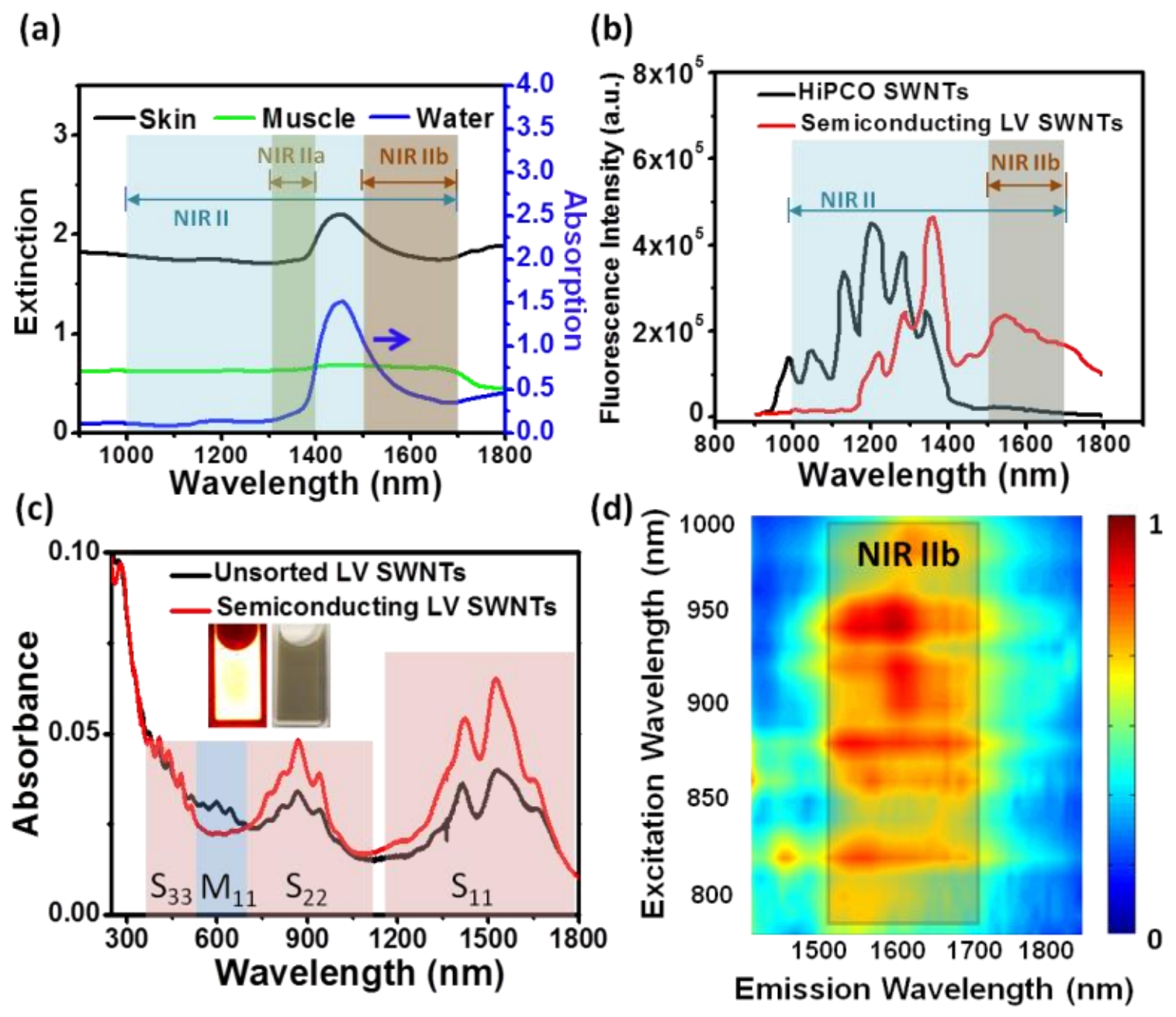

Figure 1 Semiconducting LV SWNTs for fluorescence imaging in the NIR-IIb region. a) Optical absorption spectrum of water and extinction spectra of mouse skin and mouse muscle. b) NIR fluorescence spectra of HiPCO and semiconducting LV SWNTs upon 808 nm excitation. c) Optical absorption spectra of pristine and separated semiconducting LV SWNTs. Photoluminescence (inset, left) and optical (inset, right) images of semiconducting LV SWNTs are also shown. d) The photoluminescence versus excitation (PLE) map of semiconducting LV SWNTs shows bright emission in the NIR-IIb region.

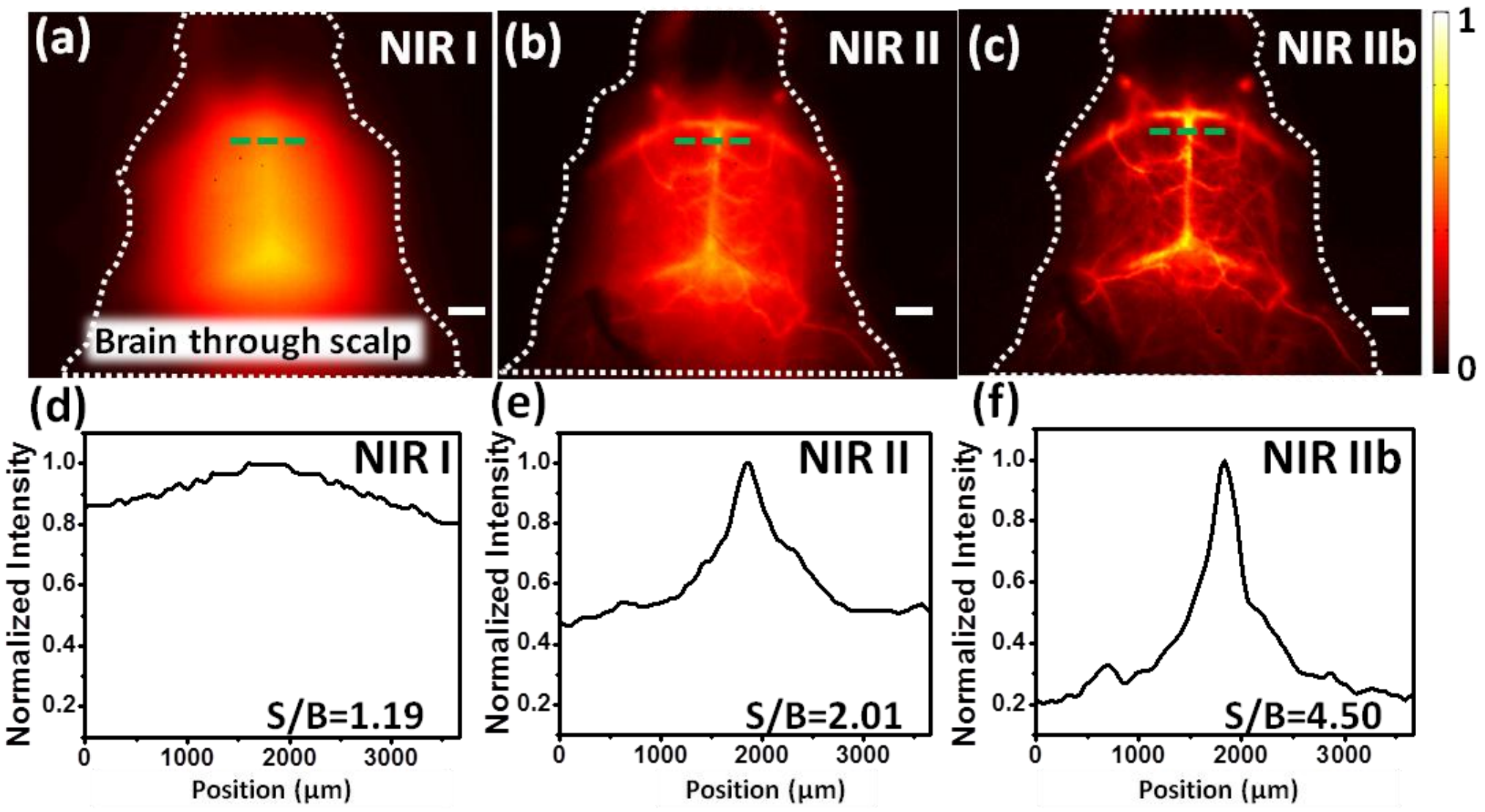

Figure 2 Fluorescence images of the cerebrovasculature of mice (*n*=2) without craniotomy in the NIR-I (a), NIR-II (b), and NIR-IIb (c) regions, with the corresponding SBR analysis shown in (d)--(f). Scale bars: 2 mm.

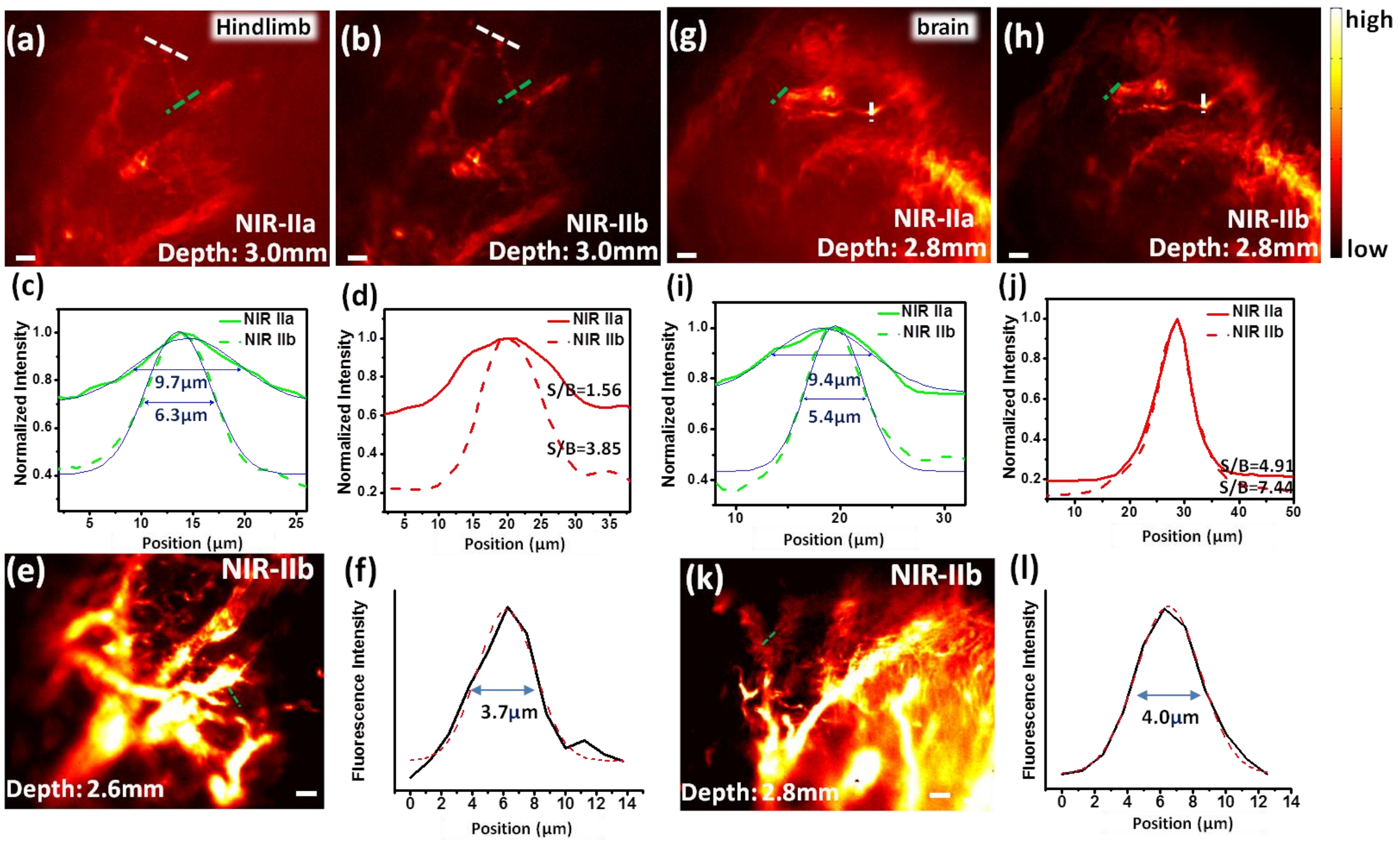

Figure 3 High-magnification vessel imaging of the hindlimb and brain of mice in the NIR-IIa and IIb windows. a,b)High-magnification microscopic images of the hindlimb vessels (*n*=2) at an imaging focal depth of 3 mm taken in the NIR-IIa and NIR-IIb windows, with the vessel FWHM width (green lines in a, b) and SBR (white lines in a, b) analysis shown in (c) and (d), respectively. The background signals for the SBR analysis were evaluated by averaging the baseline signals in the cross-sectional intensity profiles. e, f)A small blood vessel resolved by imaging in the NIR-IIb region. Similar vessel width and SBR analysis of the high-magnification cerebrovascular images in the NIR-IIa (g) and NIR-IIb (h) regions are shown in (i) and (j). k, l) A NIR-IIb fluorescence cerebrovascular image resolving a small vessel. Scale bars: 40 μm.

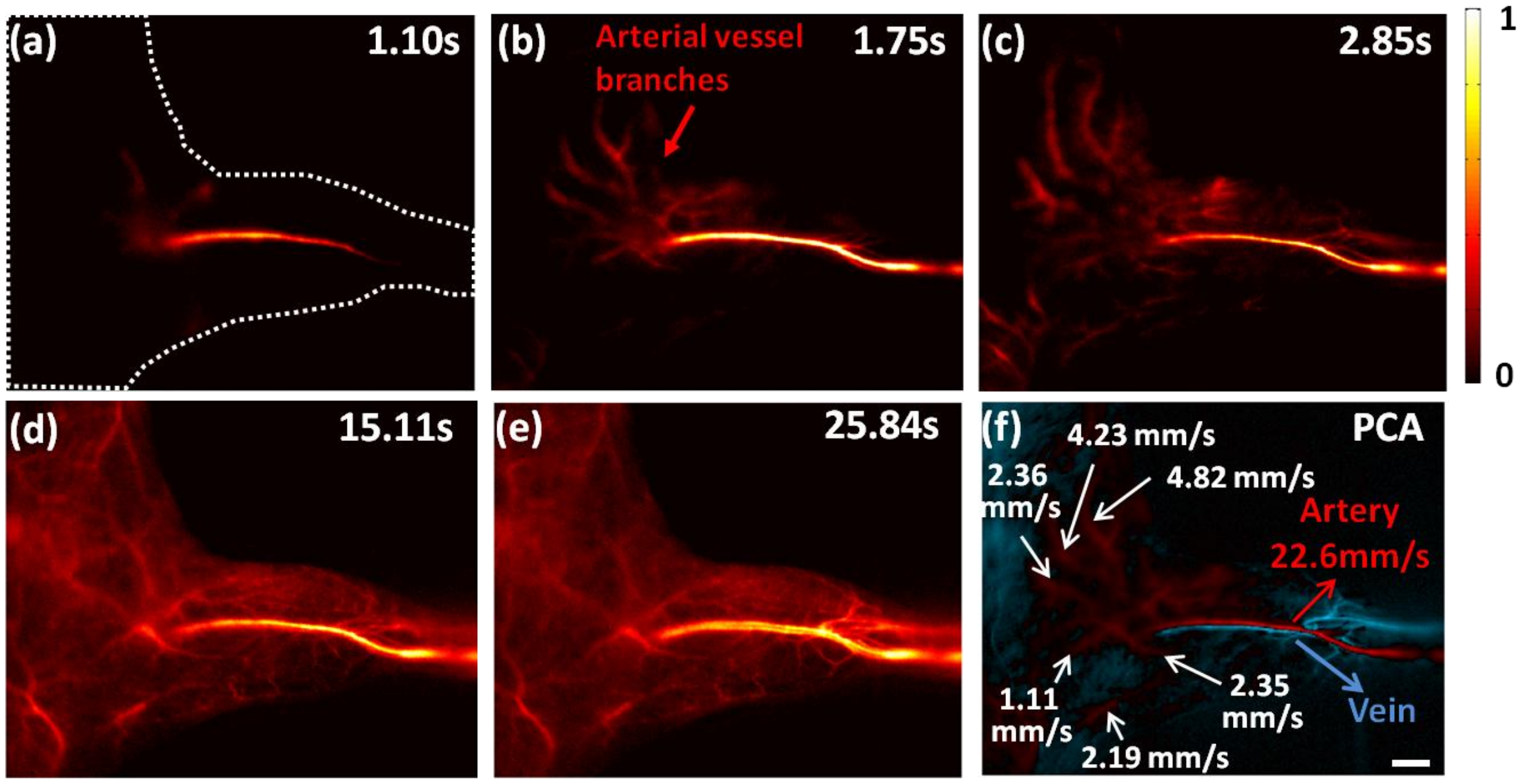

Figure 4 Video-rate NIR-IIb fluorescence imaging of mouse hindlimb vessels (*n*=3) and dynamic-contrast-based vessel type differentiation. a--e) Time-course NIR-IIb hindlimb fluorescence images after injection of semiconducting LV SWNTs into the tail vein of an athymic nude mouse. f) PCA overlaid image showing the differentiation of arterial (red) and venous (blue) vessels. The blood-flow velocities in the femoral artery and small, higher-order arterial vessels are also given. Scale bar: 2 mm.

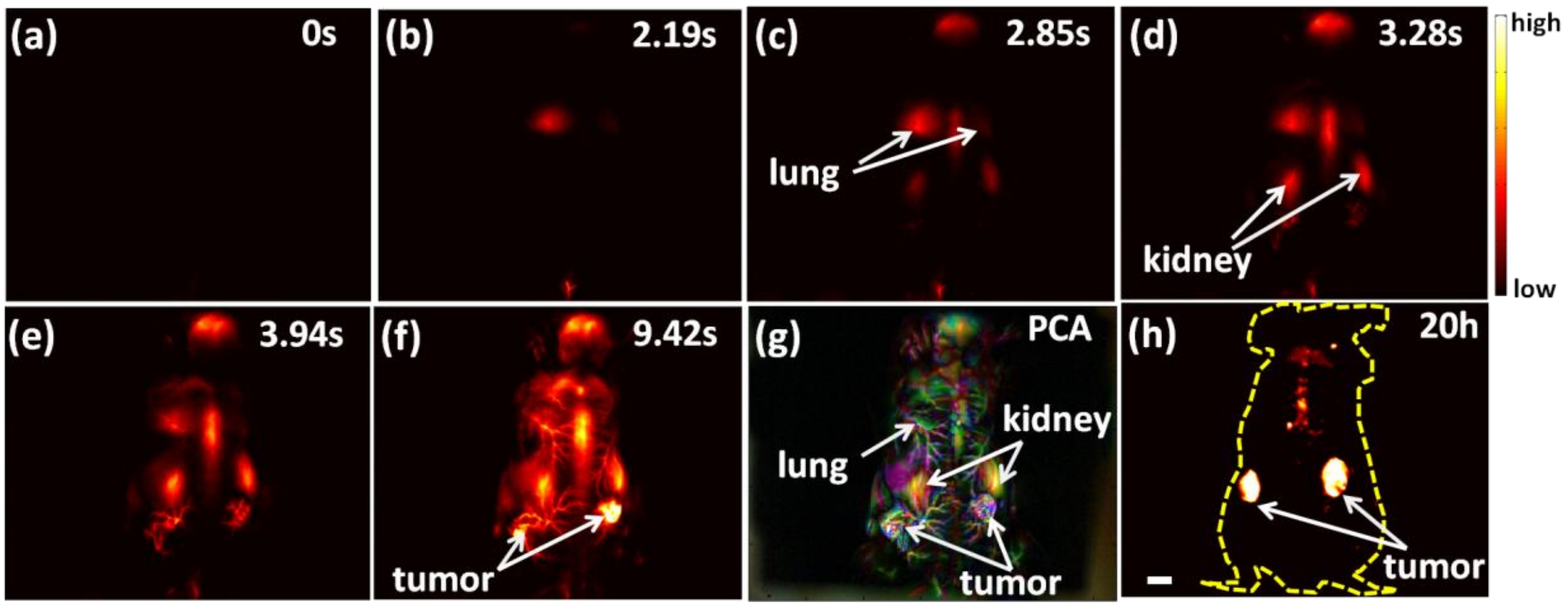

Figure 5 In vivo video-rate NIR-IIb tumor imaging in live mice (*n*=2). a--f) Time-course NIR-IIb fluorescence images of the tumor-bearing mice after intravenous injection of exchanged semiconducting LV SWNTs. g) Overlaid PCA image of the early image frames differentiating inner organs and the vascular structures surrounding tumors. h) A NIR-IIb fluorescence image of the tumor-bearing mouse taken at 20 h p.i., showing the high uptake in the tumor area that is due to the EPR effect. Scale bar: 4 mm.

# Supporting Information

## Method

**Synthesis of laser vaporization SWNTs.** Laser vaporization synthesis of SWNTs was performed by methods described previously.[1] Briefly, a 10-g solid target was prepared from graphite (Alfa Aesar, 2 – 15 μm) that was intimately mixed with 3 wt.% each nickel and cobalt powder. 10 g of the target material was pressed into a cylindrical 1 inch pellet and placed into the center of a quartz tube that was surrounded by a tube furnace for maintenance of the reaction temperature. The inert reaction environment was established by maintaining a flow of 150 standard $cm^3$/min of nitrogen gas, at a regulated pressure of 500 Torr. A Continuum PowerLite II 1064 nm Nd:YAG laser, running without Q-switching, was used to vaporize the target, and the power density was kept at ~100 W/$cm^2$. Different reaction temperatures (950, 1000 and 1125 °C) were used in separate runs to tune the diameter distribution for optimal fluorescence signal in the NIR-IIb window.

**Separation of semiconducting laser vaporization SWNTs and biocompatible functionalization.** 3 mg of laser vaporization (LV) SWNTs grown at 950 °C were sonicated in water with 1 wt.% sodium cholate (SC) for 5 h. The suspension was ultracentrifuged at 300,000g for 30 min to remove large bundles and aggregates. The supernatant was mixed with 1 wt.% sodium dodecyl sulfate (SDS) with a volume ratio of 2:3. The resulting solution was then added to a filtration column filled with ~20 ml of ally dextran-based size-exclusion gel (Sephacryl S-200, GE Healthcare).[2] A mixture surfactant solution of 0.6 wt.% SDS /0.4 wt.% SC was used to wash the column to remove the metallic nanotubes, leaving mainly semiconducting LV SWNTs in the column. The semiconducting LV SWNTs were collected by eluting the column with a solution of 1 wt.% SC. To make the SWNTs biocompatible, 1 mg/ml of (1,2-distearoyl-sn-glycero-3-phosphoethanolamine–N-[methoxy(polyethylene glycol)5000] (DSPE-mPEG(5k)) was added to the separated semiconducting LV SWNTs. After 2 min of sonication, the mixture was transferred to a 3,500 Da membrane (Fisher) for dialysis of four to five days with 5~6 water changes per day. The resulting biocompatible semiconducting LV SWNTs were then concentrated and centrifuged to remove any aggregates formed during dialysis before injection for fluorescence imaging. For video-rate NIR-IIb imaging, solutions of 200 μL of semiconducting LV SWNTs (0.43 mg/mL, 4.3 mg/kg of body weight) were used as fluorescent contrast agents.

**Ultraviolet-Vis-NIR absorbance spectrum measurements**. A Cary 6000i UV-Vis-NIR spectrophotometer was used for Ultraviolet-Vis-NIR absorbance spectrum measurements. The water absorption was measured using 1mm path cuvette (Starna Cells).

**Photoluminescence versus excitation map measurements.** Photoluminescence versus excitation map was measured by a home-built NIR spectroscopy setup upon an excitation source of a 150 W ozone-free xenon lamp (Oriel). Emission light was filtered by long-pass filters and detected using a spectrometer (Acton SP2300i) equipped with an InGaAs linear array detector (Princeton OMA-V).

**NIR fluorescence emission spectrum measurements.** The NIR fluorescence spectra were measured using the same home-built NIR fluorescence spectrometer. The excitation was provided by an 808-nm fiber-coupled diode laser (RPMC Lasers) and filtered with selected short-pass filters. The excitation light was directed to pass through a 1-mm path cuvette (Starna Cells) containing either HiPCO SWNT solution or semiconducting LV SWNT solution. The emission light was filtered using a 900-nm long-pass filter (Thorlabs) to reject the excitation light and then directed into the fluorescence spectrometer equipped with the linear InGaAs detector. All the raw emission spectra in this study were corrected for the detector sensitivity and extinction features of the filters.

**Phantom study using chicken breast or intralipid.** The excitation light was provided using the 808 nm fiber-coupled diode laser (RPMC Lasers) at a power density of 0.14 W/cm$^2$. A capillary tube was filled with a mixture of HiPCO and semiconducting LV SWNTs to afford fluorescence emission in the entire 1000-1700 nm region. The capillary tube was then inserted under a piece of chicken breast with a thickness of 4 mm and the fluorescence images of the tube were taken in NIR-II, IIa and IIb regions using a liquid-nitrogen-cooled 2D InGaAs camera (Princeton Instruments, quantum efficiency curve up to 1700 nm). Different emission filter sets were used to select various subregions (900 nm and 1000 nm long-pass filters for NIR-II region, 1000 nm, 1300 nm long-pass filters and 1400 nm short-pass filter for NIR-IIa region, 1100 nm and 1500 nm long-pass filters for NIR-IIb region). In another phantom study, a capillary tube filled with a mixture of HiPCO and semiconducting LV SWNTs was immersed in a highly scattering medium, intralipid,[3] at different depths and the images of the tube were taken in various NIR subregions (NIR-II, NIR-IIa, NIR-IIb). The same emission filter set, 2D InGaAs camera and 808 nm excitation described previously were used for this phantom study.

**Low-magnification NIR fluorescence imaging of mouse brain and hindlimb vessels.** For blood vessel imaging in brain and hindlimb, mouse was mounted on an imaging stage beneath the laser. To compare the wavelength-dependent brain imaging quality (n=2), different emission filters were used to select the collection range for NIR-I (850 nm long-pass filter and 900 nm short-pass filter), NIR-II (900 nm and 1000 nm long-pass filters), NIR-IIb regions (1100 nm and 1500 nm long-pass filters), respectively. The excitation power density is 0.14 W/cm$^2$ at 808 nm. Two achromats (200 mm and 75 mm, Thorlabs) were used to direct and focus the emission light onto the 2D InGaAs camera with 320 x 256 pixels. NIR-IIb video-rate hindlimb fluorescence images (n=3) were acquired with LabVIEW software with an exposure time of 200 ms plus a 19-ms overhead time in the readout. The emitted fluorescence from the mice was filtered using an 1100 nm long-pass filter and a 1500 nm long-pass filter to select the range of NIR-IIb region. To perform principal component analysis (PCA), early image frames immediately after injection were loaded into an array using MATLAB software.[3-4] Pixels showing up later in the video were color-coded in blue to represent venous vessels while pixels showing up earlier in the video were color-coded in red to represent arterial vessels in the overlaid PCA image.

**High-magnification microscopic imaging of hindlimb and brain vessels.** The 808 nm excitation laser and 2D InGaAs camera were the same as mentioned previously. The mouse (n=2), placed on a stereotactic stage, was imaged through a 10X objective (Bausch & Lomb) to afford a pixel size of 2.5 μm. The axial travel distance of the 10X objective, recorded by the readout of a motorized 3D translational stage, was used to determine the imaging depth after correcting for the different refractive indices of tissues.

**NIR-IIb fluorescence tumor imaging**. Balb/c mice (n=2) were shaved and inoculated with ~2 million 4T1 murine breast tumor cells for each tumor site (left and right) on the back. After ~ one week post inoculation (with an average tumor size of ~30 mm$^3$), video-rate (4.6 frames/second) NIR-IIb fluorescence tumor imaging was recorded after injection of NIR-IIb emitting semiconducting LV SWNTs using the same 808 nm excitation laser, 2D InGaAs camera and NIR-IIb region emission filters described previously (under a lower magnification than hindlimb or brain imaging). PCA was performed to group the inner organs and vessels by loading the early frames of video-rate imaging.

**Quantum yield (QY) measurement of semiconducting laser vaporization SWNTs**. The QY measurement was carried out using IR-26 in 1,2-dichloroethane (DCE) as reference (QY = ~0.1%)[5] under the excitation of an 808-nm laser.[6] A series of solutions of IR-26 in DCE with absorbance values at 808 nm to be ~ 0.10, ~ 0.08, ~ 0.06, ~0.04 and ~0.02 were prepared respectively. Optical absorbance spectra of the five solutions were measured using the method described previously (with a 1-cm path cuvette from Starna Cells). To measure the fluorescence spectra upon the 808-nm excitation (with a 1-cm path cuvette), the NIR emission light was filtered by the 900 nm long-pass filter and collected in the range of 900 to 1800 nm. Same solution preparation, absorption and fluorescence spectra measurements were performed for semiconducting laser vaporization SWNTs. All the raw emission spectra were corrected for the detector sensitivity and extinction features of the emission filter. The corrected emission spectra were then integrated in the range of 900 to 1800 nm. For both IR-26 in DCE and semiconducting laser vaporization SWNTs, the integrated fluorescence intensities of five fluorescence spectra were plotted against the actual absorbance values measured at 808 nm from optical absorbance spectra and the two slopes were obtained by linear fitting. The QY of semiconducting laser vaporization SWNTs was estimated to be ~ 0.01%, based on the following calculation:

$$QY_{SWNTs} = QY_{ref} \cdot \frac{Slope_{SWNTs}}{Slope_{ref}} \cdot \left( \frac{n_{SWNTs}}{n_{ref}} \right)^2$$

Where $QY_{SWNTs}$ is the QY of semiconducting laser vaporization SWNTs, $QY_{ref}$ is the QY of IR-26 in DCE (~0.1%), $n_{SWNTs}$ and $n_{ref}$ are the refractive indices of water and DCE.

**Mouse handling and injection**. C57Bl/6 mice (with hair on the scalp removed), athymic nude mice and balb/c mice (with hair on the back removed) were used for brain vessel imaging, hindlimb vessel imaging and tumor imaging respectively. For a typical NIR-IIb fluorescence imaging experiment, a mouse was anaesthetized by inhalation of 2.5 % isoflurane with 2 L min$^{-1}$ of oxygen and injected with 200 μL of semiconducting laser vaporization SWNTs at a concentration of 0.43 mg/mL. All procedures of mouse handling were approved by Stanford Institutional Animal Care and Use Committee (IACUC) protocols.

## ACKNOWLEDGMENT

This study was supported by grants from National Institutes of Health-National Cancer Institute (NIH-NCI) No. 5R01CA135109-02 to H.D. LV SWNTs synthesis was performed at NREL, and was supported by the Solar Photochemistry Program, Division of Chemical Sciences, Geosciences, and Biosciences, Office of Basic Energy Sciences, U.S. Department of Energy (DOE), Grant DE-AC36-08GO28308 (for developing carbon nanomaterial with advanced properties and spectroscopic characterizations).

# Supplementary Figures

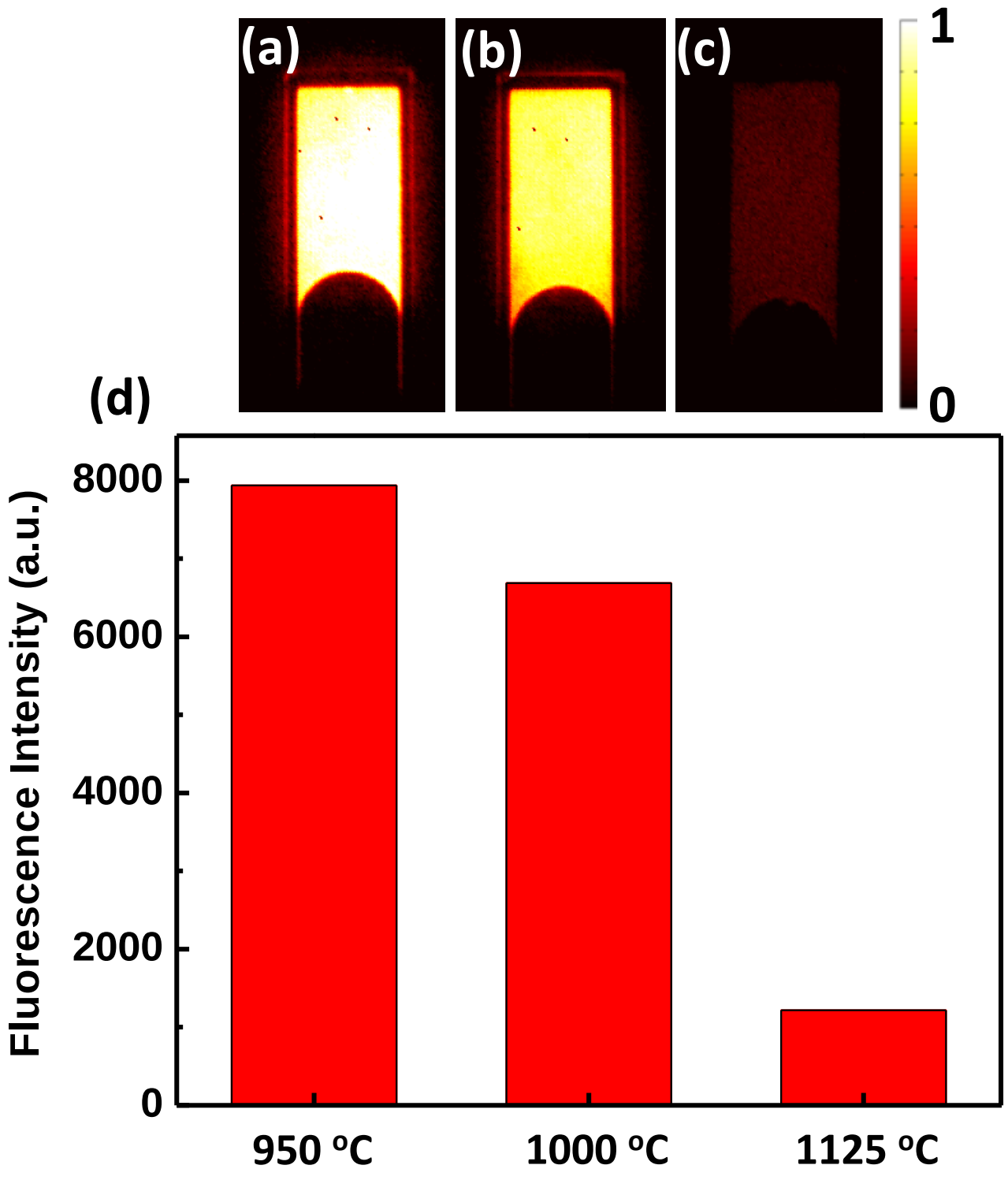

**Figure S1.** NIR-IIb fluorescence brightness comparison of laser vaporization SWNTs (LV SWNTs) grown at 950 ˚C (a), 1000 ˚C(b) and 1125 ˚C (c) upon an 808 nm excitation on the per mass basis[2a] is shown in (d).

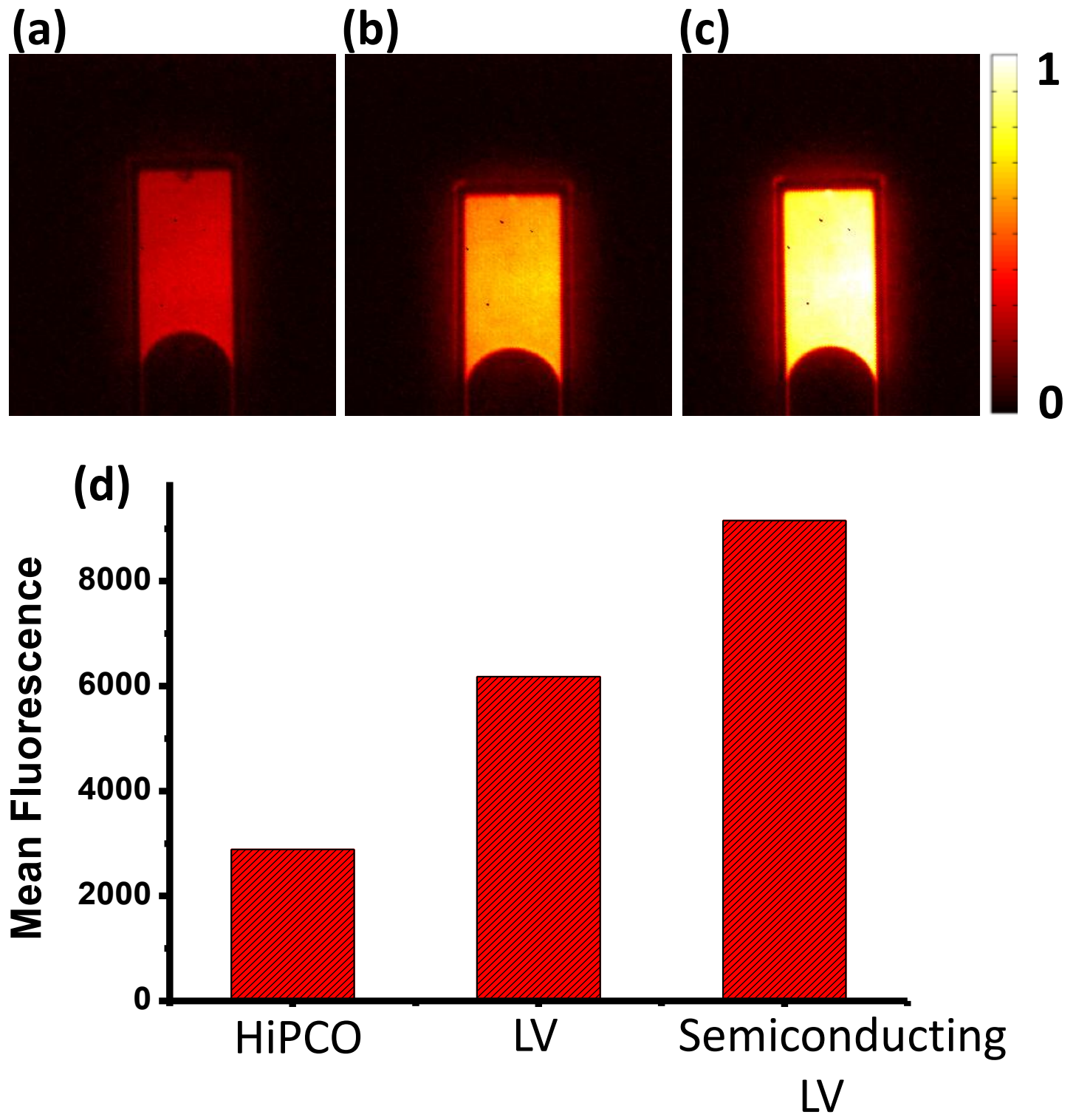

**Figure S2.** NIR-IIb brightness comparison of HiPCO SWNTs (a), pristine LV SWNTs (b) and separated semiconducting LV SWNTs (c) on the per mass basis.[2a] Compared to the HiPCO SWNTs, pristine LV nanotubes grown at a synthesis temperature of 950 ℃ exhibited two-fold higher brightness in NIR-IIb region on the per mass basis, and the separated semiconducting LV SWNTs showed three times higher brightness in NIR-IIb region (d).

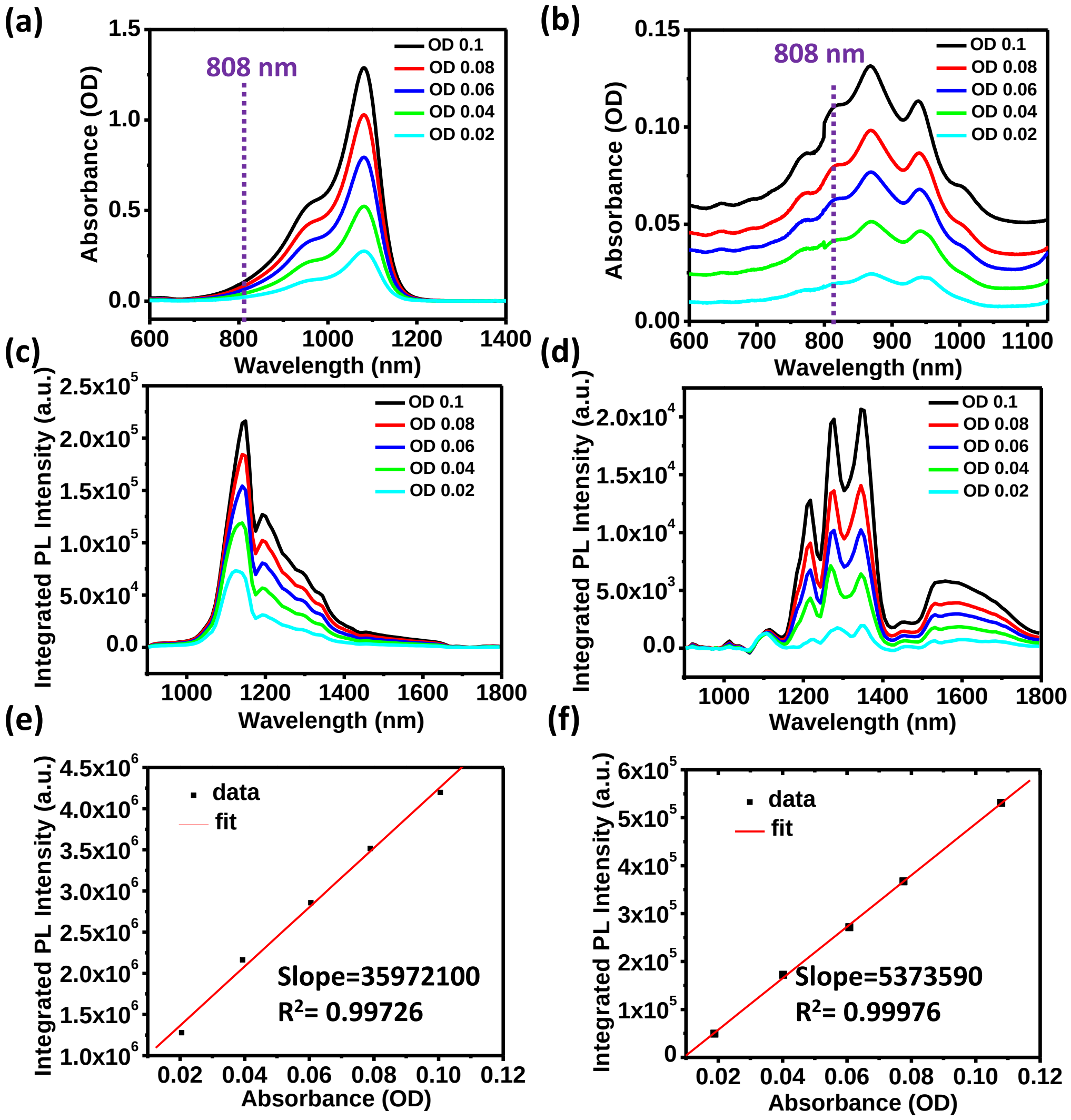

**Figure S3.** Quantum yield measurement of semiconducting laser vaporization (LV) SWNTs using IR-26 in DCE (quantum yield of ~0.1%) as reference.[5] The reported quantum yield values of IR-26 in DCE are not consistent in the literature, with variations up to an order of magnitude.[7] Here, we chose the IR-26 quantum yield value from a recent study,[5] which carefully discussed the important factors that may affect the quantum yield measurement. (a) Optical absorbance spectra of a series of five solutions of IR-26 in DCE with absorbance values at 808 nm to be ~ 0.10, ~ 0.08, ~ 0.06, ~0.04 and ~0.02. (b) Optical absorbance spectra of a series of five semiconducting LV SWNTs solutions with absorbance values at 808 nm to be ~ 0.10, ~ 0.08, ~ 0.06, ~0.04 and ~0.02. (c) NIR fluorescence spectra of the five IR-26 solutions in the range of 900-1800 nm under an excitation of 808 nm laser. (d) NIR fluorescence spectra of the five semiconducting LV SWNTs solutions in the range of 900-1800 nm under an excitation of 808 nm laser. (e) The integrated emission intensities of five fluorescence spectra of IR-26 solutions plotted against the actual absorbances at 808 nm. The slope was obtained by linear fitting. (f) The integrated emission intensities of five fluorescence spectra of semiconducting LV SWNTs solutions plotted against the actual absorbances at 808 nm. The slope was obtained by linear fitting. The QY of semiconducting LV SWNTs was found to be ~0.01%. Although this QY was low, the LV SWNTs were found sufficiently bright for *in vivo* biological imaging in NIR-IIb region under an 808-nm laser excitation power density of 0.14 $W/cm^2$ (well below the safe exposure limit of 0.33 $W/cm^2$ [8]). The detailed measurements and calculation are shown in Method section.

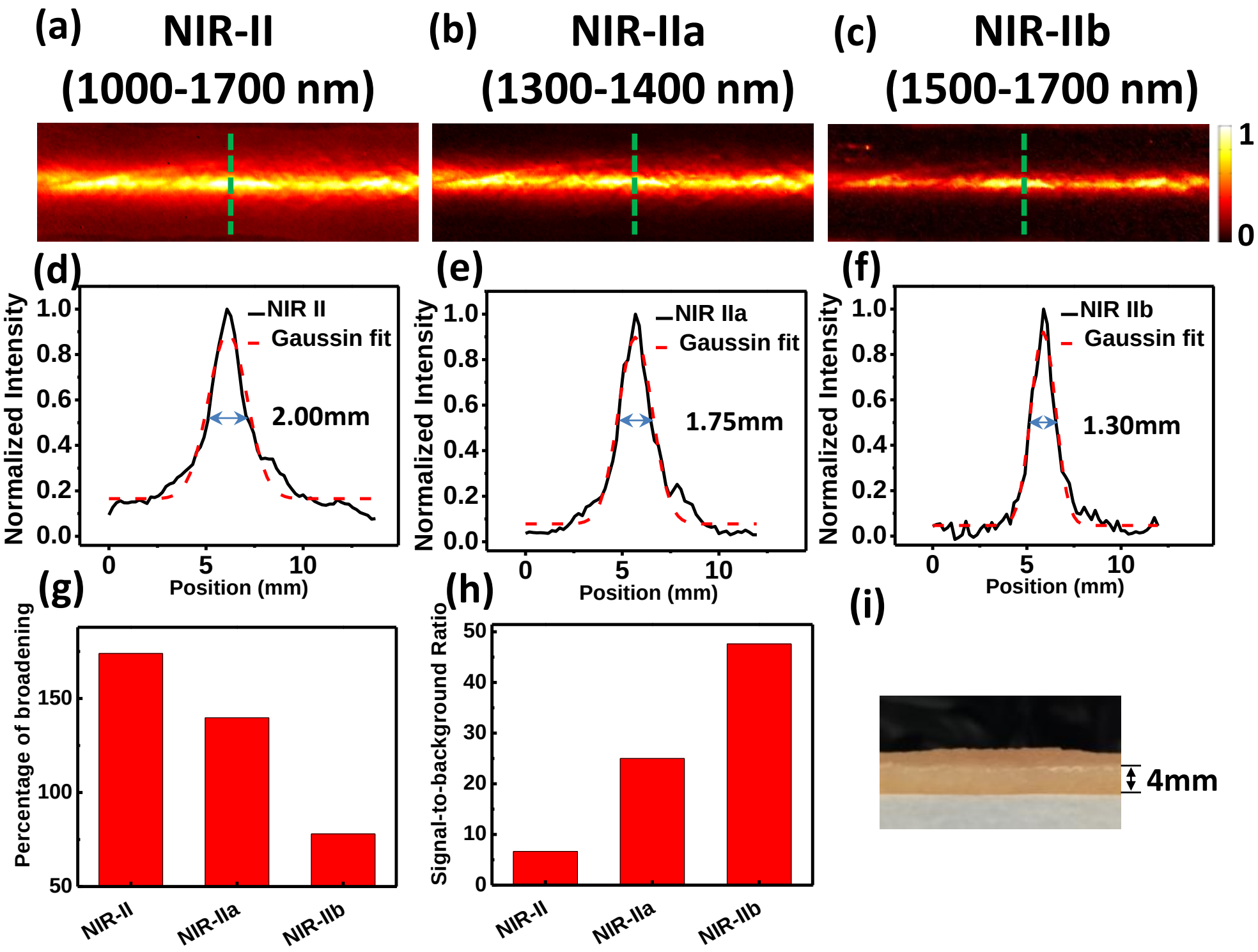

**Figure S4.** To compare the photon scattering effects in NIR-II (1000-1700 nm), NIR-IIa (1300-1400 nm) and NIR-IIb (1500-1700 nm) regions, we performed a mock tissue phantom study by inserting a capillary tube filled with a mixture of HiPCO and semiconducting LV SWNTs under a piece of chicken breast with a thickness of ~ 4 mm. Due to the higher scattering of shorter wavelength photons by chicken tissues, images acquired in the NIR-II and NIR-IIa windows were more blurry than in NIR-IIb (a-c), with the apparent capillary tube width much wider compared to the tube width measured in the NIR-IIb window (2d-f). By measuring the Gaussian-fitted full width at half maximum (FWHM) of the cross-sectional intensity profiles, the apparent widths of the capillary tube imaged in NIR-II, IIa and IIb region were found to be broadened by 173%, 140% and 78%, respectively (g). For each NIR subregion, the percentage of broadening is calculated by dividing the FWHM width of the tube imaged under chicken breast (2.00, 1.75, 1.30 mm) by the FWHM width of the same tube imaged without chicken breast (0.73 mm) minus 100%. SBR analysis of the images is shown in h. Figure i shows a photo of the 4-mm thick chicken breast.

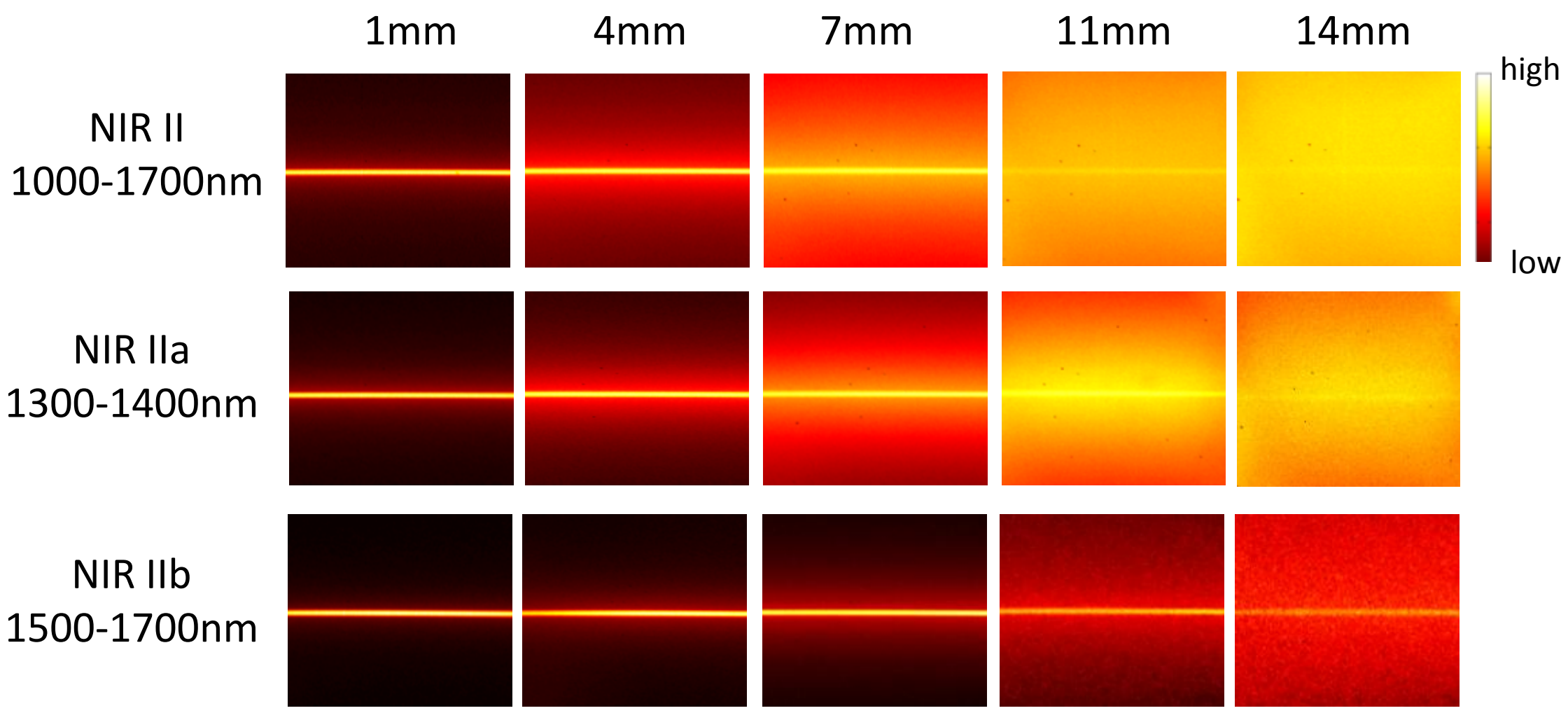

**Figure S5.** Photon scattering phantom study using intralipid. A capillary tube filled with a mixture of HiPCO and semiconducting LV SWNTs was immersed in a scattering medium, intralipid[3] at various depths from 1 mm to 14 mm. Fluorescence images of the capillary tube were taken in NIR-II, IIa and IIb regions. Due to the much reduced scattering of photons, the tubes were clearly resolved in the images in NIR-IIb region with a clear and sharp resolution up to a penetration depth of 14 mm. In contrast, images taken in NIR-II and IIa regions became blurry with much loss of clarity, especially at depths > 7 mm.

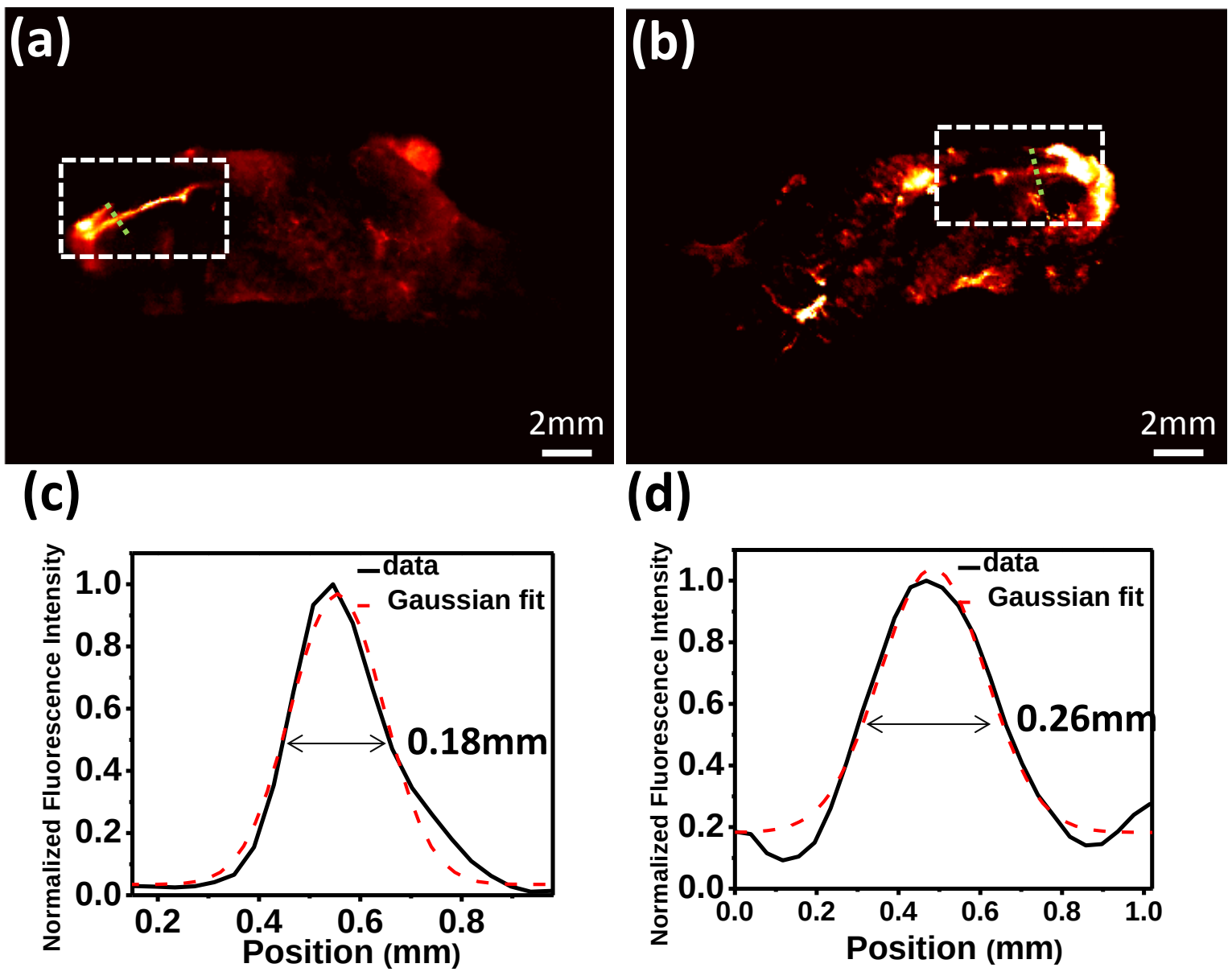

**Figure S6**. To evaluate the accuracy of our vessel size measurements in the mouse, we studied light scattering effects in NIR-IIb region by intravenously injecting semiconducting LV SWNTs to a mouse tail vein and slicing a piece of ~2.5 mm thick hindlimb tissue from the mouse after dissection. We imaged a blood vessel (at the tissue surface) using a low-magnification mode in the NIR-IIb region (a, pixel size ~78 μm). Then the whole hindlimb piece was flipped over and the same blood vessel was imaged through the 2.5 mm thick tissue (b, please note that a and b are bilateral symmetric). A broadening of ~44% in vessel size was measured due to the light scattering caused by the ~2.5 mm thick tissue (c-d).

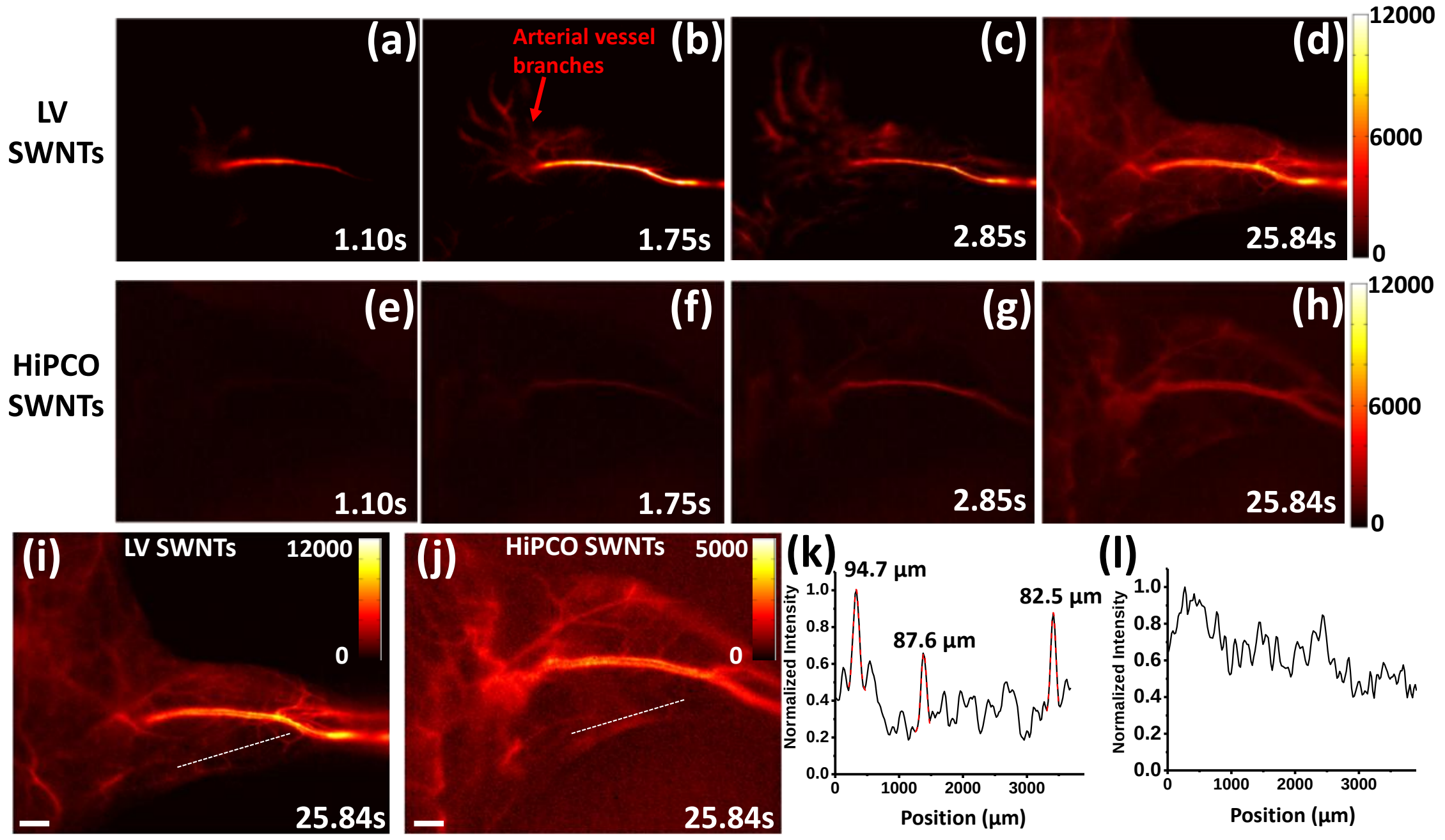

**Figure S7**. Mouse hindlimb imaging in the 1500-1700 nm NIR-IIb region using HiPCO and LV SWNTs. (a-d) Time-course images of video-rate hindlimb fluorescence imaging in the NIR-IIb region using semiconducting LV SWNTs as emitters (data also shown in the Figure 4 in the main text). (e-h) Time-course images of video-rate hindlimb fluorescence imaging in the NIR-IIb region using HiPCO SWNTs as emitters. (i-j) Rescaled d and h (images taken at ~25.8 s post injection of LV SWNTs and HiPCO SWNTs) based on the maximum fluorescence intensity in each image. The average fluorescence intensity of femoral vessels in i is ~ three times higher than the one measured in j, suggesting the benefit of using semiconducting LV SWNTs with larger average diameter as NIR-IIb emitters. (k-l) Cross-sectional intensity profiles of three small vessels marked in i and j. The apparent widths of these vessels were measured to be 94.7, 87.6, 82.5 µm, respectively by full width at half maximum (FWHM) in k using semiconducting LV SWNTs. However, these vessels were not well resolved in l when using HiPCO SWNTs as emitters, due to the low signal to background ratios caused by the lack of large diameter nanotubes emitting in the 1500-1700 nm NIR-IIb region. Here imaging conditions for HiPCO SWNTs and semiconducting LV SWNTs, such as SWNT injection dose, laser power density, imaging exposure time and frame rate, 2D InGaAs detector and emission filters, are the same.

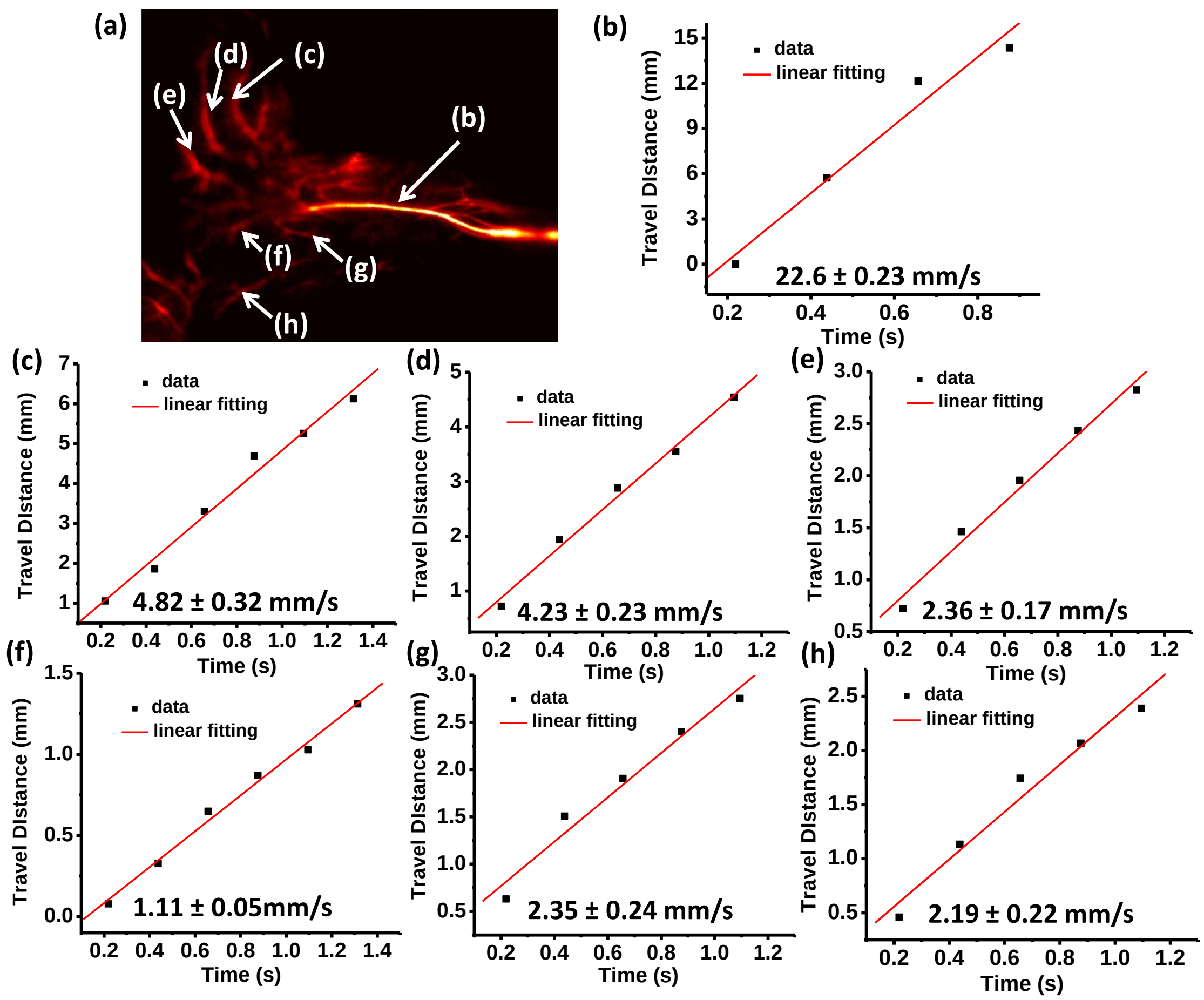

**Figure S8**. Blood flow speed quantification in various hindlimb arterial vessels based on real-time NIR-IIb hindlimb fluorescence imaging. (a) A NIR-IIb fluorescence image of hindlimb vessels after intravenously injecting a solution of semiconducting LV SWNTs. The blood flow velocities of the femoral artery (labeled b) and deeper, higher-order arterial branches (labeled c-h) are quantified in Figure b and c-h respectively. (b-h) Blood flow velocity quantification by plotting the distance travelled by blood front against time. In each plot, the flow distance shows a linear increase versus time, exhibiting blood flow speeds of 22.6 mm/s in the femoral artery and 1.1-4.8 mm/s in the smaller arterial branches.

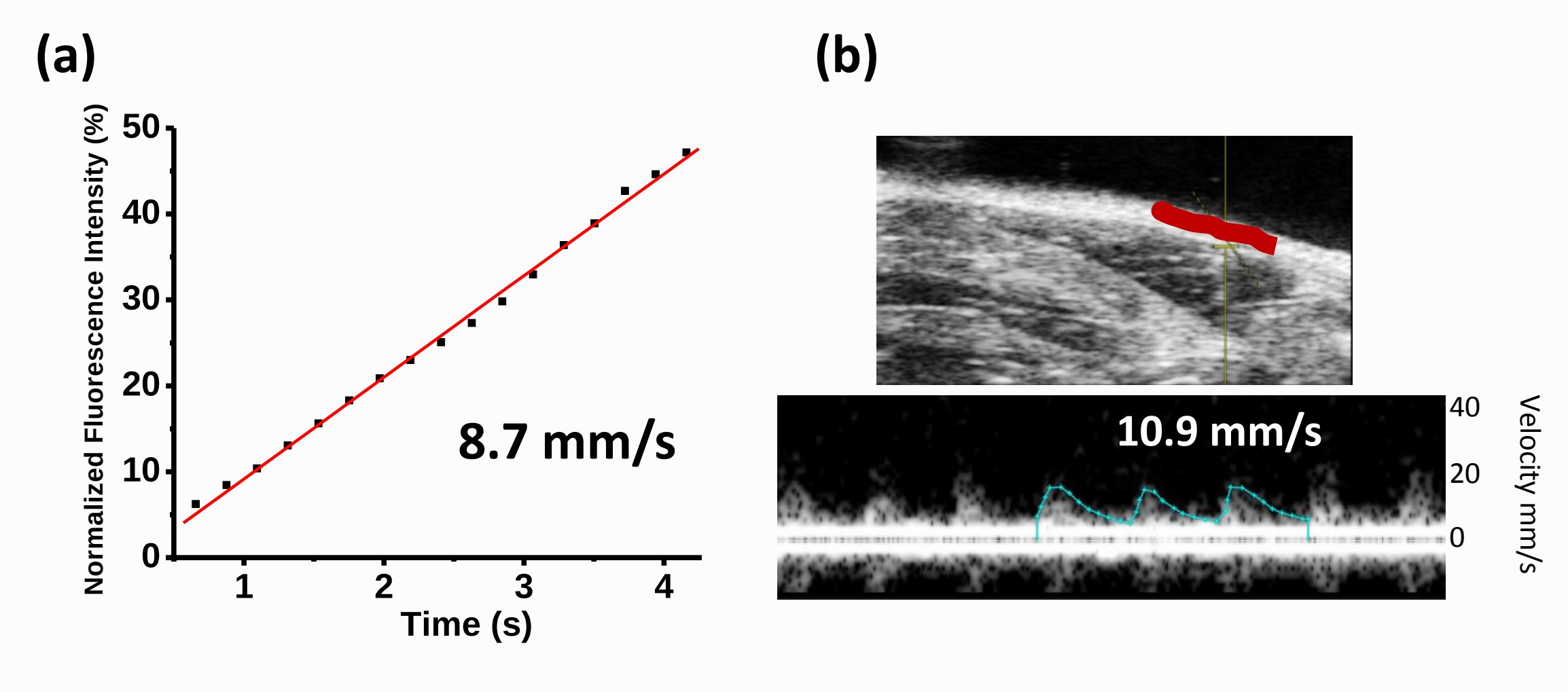

**Figure S9.** To verify the accuracy of our fluorescence-based velocity measurements, we compared the femoral vessel blood velocities of another mouse measured by both video-rate NIR-IIb fluorescence imaging method (a) and ultrasound method (b) side by side.[4] Ultrasound method is known to be the gold standard for *in vivo* blood flowmetry. We found that the Doppler-derived velocity measured by ultrasound method agreed well with the one measured by video-rate fluorescence imaging method (calculated based on the method described in a previous study[4]) with a deviation of ~ 20 %. We attribute this deviation to the slight change in mouse body temperature caused by different levels of anaesthesia during fluorescence imaging and ultrasound imaging. Meanwhile, we found that the blood velocities in the smaller arterial branches were hard to measure using ultrasound method due to the small dynamic range and low spatial resolution, showing the advantageous of NIR-IIb fluorescence imaging by providing a broader blood velocity dynamic range with high resolution.

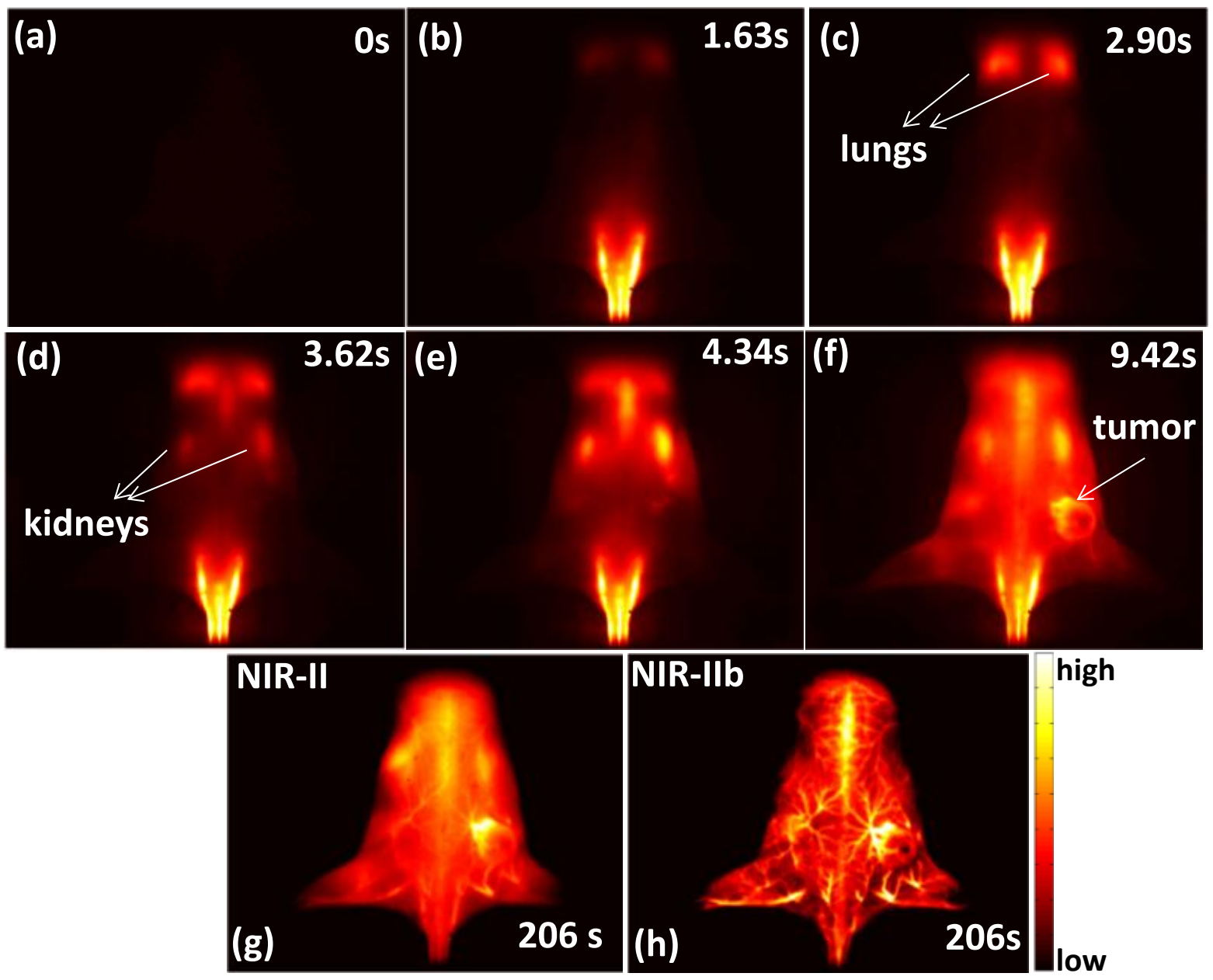

**Figure S10**. To compare the wavelength-dependent fluorescence-based tumor imaging, we performed tumor imaging in both 1000-1700 nm NIR-II region[9] and 1500-1700 nm NIR-IIb region on the same group of mice (n=2) side by side using HiPCO SWNTs and semiconducting LV SWNTs as emitters respectively. We first performed video-rate tumor imaging in the NIR-II region on a tumor bearing balb/c mouse by intravenously injecting 200 μL of biocompatible HiPCO SWNT solution[9] (with confined emission in the range of 1000-1400 nm) and the time-course images are shown in a-f. Although the major organs (such as lungs and kidneys) showed up with a modest imaging contrast shortly after injection, the vessels structure surrounding the tumor area and the rest of mouse body, which appeared prominently when imaged at video rate in the NIR-IIb region using semiconducting LV SWNTs (Figure 5 in the main text), exhibited greater loss of clarity in the NIR-II region due to the much increased light scattering from photons shorten than 1400 nm that travel through mouse tissues. To further compare wavelength-dependent tumor imaging on the same mouse side by side, we intravenously injected another 200 μL of semiconducting LV SWNT solution into the same mouse as NIR-IIb emitters and compared the tumor-bearing mouse in both NIR-II (g) and NIR-IIb (h) regions side by side. Quantitatively, the number of vessels resolved in the NIR-IIb region was found to much higher than the ones resolved in NIR-II, demonstrating of the benefit of higher imaging resolution with deep penetration due to the reduced light scattering in this long wavelength NIR-IIb region. Therefore although imaging of tumors and inner organs is achievable in shorter wavelength regions such as NIR-II region (as shown in both Figure S10 and a previous paper by our group[9]), imaging in the NIR-IIb region is advantageous due to the reduced light scattering, resolving fine features with higher resolution and deeper tissue penetration, beneficial for biomedical applications.

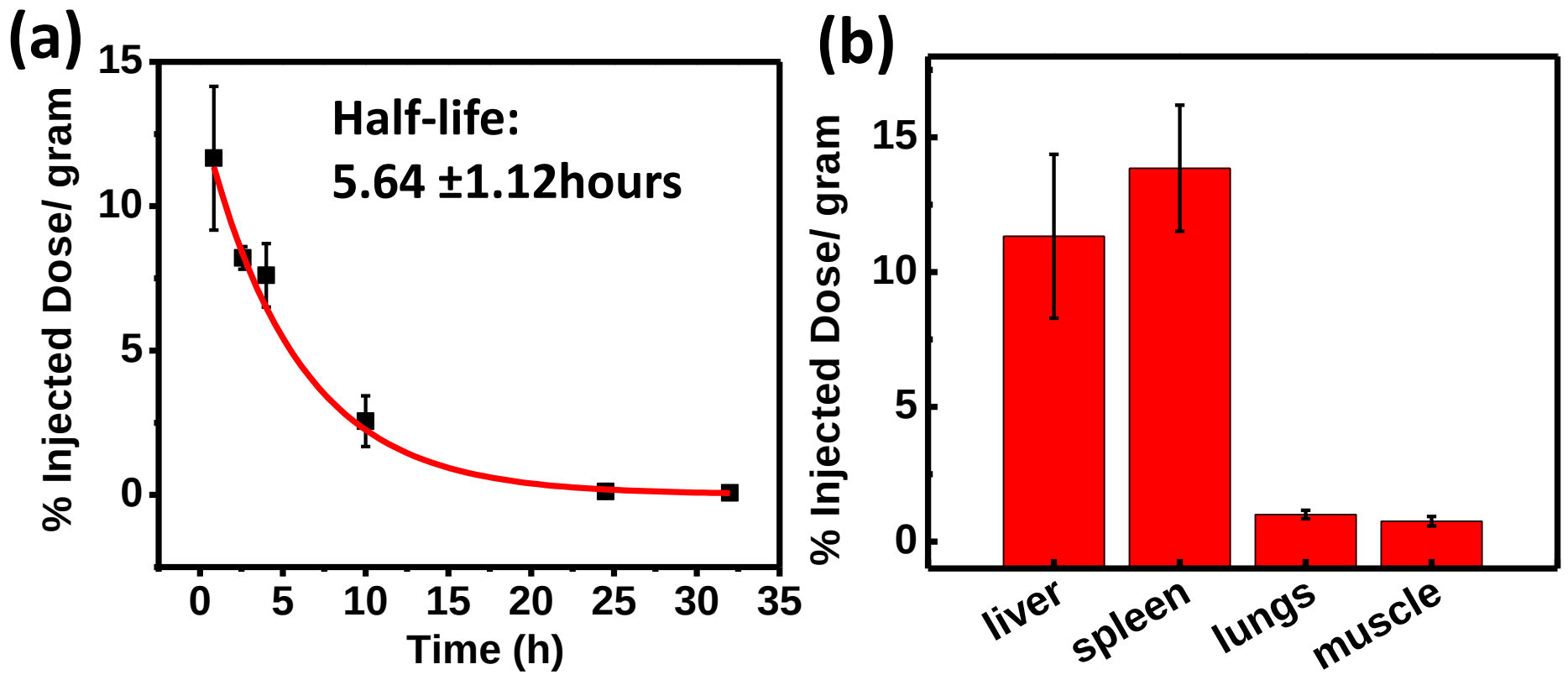

**Figure S11**. To study the clearance of SWNTs from mouse blood circulatory system, we performed an experiment by intravenously injecting semiconducting LV SWNTs to a group of balb/c mice (n=2) and collecting blood at various time points post injection (up to 30 hours).[9-10] By measuring SWNT concentration in each blood sample (mixed with tissue lysis buffer) based on the fluorescence intensity in the NIR-IIb region, we estimated the half-life of SWNT blood circulation to be ~ 5.6 hours (a). We further studied the biodistribution of semiconducting LV SWNTs in mice.[9-10] At 7 days post injection, the mice were sacrificed and all major organs were collected. Each organ was weighed before dissolving in 1 ml of Solvable with 2% sodium cholate solution. The resulting solutions were heating at 80 ˚C for overnight to fully dissolve the organs. The fluorescence intensity of each solution was measured and the SWNT uptake in each organ was calculated based on the fluorescence intensity of the injected solution (after dilution). The semiconducting LV SWNTs were found to be largely accumulated in liver (11 percent injected dose per gram, or 11% ID/gram) and spleen (14% ID/gram) of the reticuloendothelial system (RES) (b). The uptakes in other organs (such as lungs and muscles) were found to be about or less than 1% ID/gram. Autofluorescence from control mouse organs and blood was found to be extremely low in the long wavelength NIR-IIb window,[11] which makes sure that the fluorescence from each organ/blood sample nearly all comes from SWNTs. The concentration of SWNTs in all organ and blood samples measured fell well within the linear range where the fluorescence intensity is proportional to the SWNT concentration.